\documentclass[
    ,final            
  ]
  {aipproc}

\layoutstyle{6x9}

\usepackage{aas_macros}
\usepackage{ dsfont }
\usepackage{ esint }
\usepackage{graphicx}
\newcommand{\wh}{\;\widehat{=}\;}
\newlength{\fgwidth}

\begin{document}

\title{Information field theory}


\author{Torsten En{\ss}lin}{
  address={Max Planck Institute for Astrophysics\\
 Karl-Schwarzschildstr. 1, 85741 Garching bei M{\"u}nchen, Germany\\
 \url{http://www.mpa-garching.mpg.de/ift}}
}

\begin{abstract}
Non-linear image reconstruction and signal analysis deal with complex inverse problems. 
To tackle such problems in a systematic way, I present information field theory (IFT
) as a means of Bayesian, data based
inference on spatially distributed signal fields. IFT is a statistical field
theory, which permits the construction of optimal signal recovery
algorithms even for non-linear and non-Gaussian signal inference
problems. IFT algorithms exploit spatial correlations of the signal fields and benefit from techniques developed to investigate quantum and statistical field theories, such as Feynman diagrams,
re-normalisation calculations,
and thermodynamic potentials. 
The theory can be used in many areas, and applications in cosmology and numerics
are presented. 
\end{abstract}

\maketitle


\section{Information field theory}
\subsection{Field inference}

A physical field is a function over some continuous space. The air temperature over Europe, the magnetic field within the Milky Way, or the dark matter density in the Universe are all fields we might want to know as accurately as possible. 
Fortunately, we have measurement devices delivering us data on these fields. But the data is always finite in size, whereas any field has an infinite number of degrees of freedom, the field values at all locations of the continuous space the field is living in.  Since it is impossible to determine an inifinte number of unknowns from a finite number of constraints, an exact field reconstruction from the data alone is impossible. Additional information\footnote{\textit{Information} is  understood here in its original and colloquial meaning \textit{to give form to the mind}, or ``Information is whatever forces a change of rational beliefs'' \cite{2007AIPC..954...11C}. Mathematically, information theory is just probability theory. In some contexts, but not here, negative entropy is called \textit{information} as well, although it is rather a measure of the amount of information than information itself.} is needed.

Additional information might be available in form of physical laws, statistical symmetries, or smoothness properties known to be obeyed by the field. A unique field reconstruction might still be impossible, but the configuration space of possible field realizatoins might be sufficently constrained to single out a good guess for the field. 

The  combination of data and additional information is preferentially done in an information theoretically correct way by using probabilistic logic. Information field theory (IFT) is therefore information theory applied to fields, Bayesian reasoning with an infinite number of unkowns \cite{Lemm2003, 2009PhRvD..80j5005E}.  For a physicists, it is just a statistical field theory, as we will see, and can borrow many concepts and techniques developed for such. Mathematically, it deals with stochastic functions and processes and benefits from the theory of Gauss-, Markov-, L\'evy-, and other random processes.

The main difference of IFT to the usual Bayesian inference is that the continuity of the physical space plays a special role. The fact that many physical fields do not exhibit abitrary roughness due to their causal origins implies that 
field values at nearby locations are similar, and typically more so the closer the locations are. The consequent exploitation of any knowledge on the field correlation structure permits us to overcome the ill-posedness of the field reconstruction problem. 

\subsection{Path integrals}
\label{sec:path_integral}

Probabilistic reasoning requires that probability density functions (PDFs) can properly be defined over the space of all possibilities \cite{2003prth.book.....J}. The configuration space of a field is of infinite dimensionality, since every location in space carries a field degree of freedom. A little bit of thought is therefore needed on how to deal with PDFs over functional spaces before we can use probabilistic logic for field inference.

Let  $s=(s_x)_x$ be our unknown signal field living on some physical space $\Omega = \{ x\}_x$, e.g. $s$ might be a real- or complex-valued function $s:\Omega \rightarrow \mathds{R}$ or $\mathds{C}$.
The configuration space of $s$ could be constructed if the set of  physical locations in space would be finite, say of size $\mathcal{N}$ with $\Omega=\{x_1, \,\ldots,\,x_\mathcal{N}\}$.  Then the field values at these locations would form a finite-dimensional vector $s=(s_{x_1}, \,\ldots, \, s_{x_\mathcal{N}}) \equiv (s_i)_{i=1}^{\mathcal{N}}$ and the configuration space would be just the space of such vectors. We could then define any PDF on this vector space, like a signal prior $\mathcal{P}(s)$. This would also permit us to calculate configuration space integrals, like the signal prior expectation value of any function $f(s)$ of the discretized signal
\begin{equation}
 \langle f(s) \rangle_{(s)} \equiv \int\! \mathcal{D}s\, f(s)\, \mathcal{P}(s) \equiv \left( \prod_{i=1}^\mathcal{N}\,\int \! ds_i\, \right) \, f(s)\, \mathcal{P}(s).
\end{equation}
Now, we just have to require that the continuous limit of this discretization is possible yielding a path integral. This requires on the one hand that our space discretization gets finer everywhere with $\mathcal{N}\rightarrow \infty$ and on the other hand that all the involved quantities ($s$, $f(s)$, $ \mathcal{P}(s)$) behave well under this limit. The latter just implies that any reasonable expectation value $\langle f(s) \rangle_{(s)}$ should not depend on the the discretization resolution if the resolution is chosen sufficiently high. Thus, the definitions of the quantities  $s$, $f(s)$, and $ \mathcal{P}(s)$ cannot depend on any grid specific properties and must be possible in the contiuum limit. We turn the last requirement into a design property:

\vspace{1em}
\noindent%
\framebox{\begin{minipage}[t]{1\columnwidth}%
\begin{center}%
\vspace{1em}
\textbf{An information field theory is defined over continuous space.}\end{center}\label{def:IFT}%
\end{minipage}}
\vspace{1em}

Space discretization can be done in a second step, if needed in order to do inference on a computer\footnote{A code to handle this discretization properly is {\bf \textsc{NIFTy}} -- {\bf N}umerical {\bf I}nformation {\bf F}ield {\bf T}heor{\bf y}.}. However, the theory shall not contain any discretization specific element. This distinguishes IFT from many other proposed methods for field inference, Bayesian or not, since these often have definitions tightly linked to specific space discretizations, e.g. by using concepts like pixel statistics and nearest pixel field differences. The inference results of such methods might depend on the chosen space discretization and might not be resolution independent. For IFT, we require that given a sufficiently high spatial resolution, the solution shall not change significantly with further resolution increase or with a rotation of the computational grid.

Dealing with an infinite number of degrees of freedom, we should not be surpriesed about mathematical objects in IFT that are infinite (e.g. configuration space volumes, entropies) or zero (e.g. properly normalized field PDFs) in the continuous limit. As long as the quantity we are interested in is well defined in the continuous limit (i.e. posterior mean field), we should not worry too much, since divergences of auxilliary quantities are well known in field theory and usually harmless. Frequently, only the well behaved differences or ratios of such unbound objects are of actual interest (relative entropies, energy differences). 

It is most instructive to see how IFT works in a concrete example. We therefore turn now to the simplest possible case.

\subsection{Free theory}
\subsubsection{Information Hamiltonian}

Suppose we are interested in a zero mean random field $s$, our signal,  over continuous $u$-dimensional Euclidean space $\Omega =\mathds{R}^u$. The a priori field knowledge might be that the field is following homogeneous and isotropic Gaussian statistics,
\begin{equation}
 \mathcal{P}(s)=\mathcal{G}(s,S)=\frac{1}{\sqrt{|2\pi S|}}\,\exp\left(-\frac{1}{2}s^\dagger S^{-1}s\right),
\end{equation}
with the field covariance matrix $S=\langle s\,s^\dagger\rangle_{(s)}$ being known if the field power spectrum is known from some physical considerations. E.g., the field might be the cosmic density field for which, given a cosmological model, the power spectrum can be calculated theoretically. The field $s$ is here regarded as a vector from a function vector space (the configuration space of $s$) with the scalar product 
\begin{equation}
 s^\dagger j = \int_\Omega \!\! dx\;\overline{s_x} j_x .
\end{equation}
The determinant $|S|$ is of course poorly defined in the continuum limit, but it is a perfectly sensible quantity in any finite space discretization. Since we only use $|S|$  to ensure proper normalization of $ \mathcal{P}(s)$, whereas our interest is in inferring $s$, there is nothing to worry about. 

Our measured data set $d=(d_i)_i=(d_1, \,d_2, \, \ldots)$ enters the game via a data model. In the simplest case of a linear measurement, the data is
\begin{equation}
 d=R\,s + n
\end{equation}
with $R\,s= \int dx\,R_{i\,x}\,s_x$ being the signal response and $n=(n_i)_i=(n_1, \, \ldots)$ being the noise. 
The response operator $R$ encodes the point spread function of our instrument, the scanning strategy of the used telescope, and any (linear) operation done on the data, like a Fourier transformation in case we measure with an interferometer.
The noise shall here also obey Gaussian zero mean statistics with known covariance $N= \langle n\,n^\dagger \rangle_{(n)}$ (now with the data space scalar product $n^\dagger d=\sum_i \overline{n_i} d_i$) so that the data likelihood given the signal is
\begin{equation}
 \mathcal{P}(d|s)=\mathcal{G}(d-Rs,N).
\end{equation}

Now the signal field posterior can be constructed via Bayes theorem,
\begin{equation}
 \mathcal{P}(s|d)=\frac{\mathcal{P}(d|s)\,\mathcal{P}(s)}{\mathcal{P}(d)}\equiv \frac{e^{-H(d,s)}}{Z_d},
\end{equation}
where we just defined the information Hamiltonian and its partition function,
\begin{eqnarray}
 H(d,s)&\equiv& -\ln \mathcal{P}(d,s)= -\ln \mathcal{P}(d|s)-\ln \mathcal{P}(s)\;\mbox{and}\\
 Z_d &\equiv& \int \mathcal{D}s\, e^{-H(d,s)} =  \int \mathcal{D}s\, \mathcal{P}(d,s)=\mathcal{P}(d),
\end{eqnarray}
in order to translate Bayesian language into that of statistical  field theory. Thus, we can use any technique developed  for such in order to do our signal inference.
\subsubsection{Wiener filter}
For our specific linear and Gaussian measurement problem, the Hamiltonian
\begin{equation}
 H(d,s)\wh \frac{1}{2}\, (d-Rs)^\dagger N^{-1}(d-Rs)+\frac{1}{2}\, s^\dagger S^{-1} s
\end{equation}
is quadratic in $s$. We have dropped here irrelevant $s$-independent terms, as indicated by  ``$ \wh $''. This Hamiltonian can be brought into the canonical form
\begin{equation}
 H(d,s)\wh  \frac{1}{2}\, (s-m)^\dagger D^{-1}(s-m)
\end{equation}
via  quadratic completion, where $ m = D\,j$, $D =(S^{-1}+R^\dagger N^{-1} R)^{-1}$, and $j   =R^\dagger N^{-1} d$.
This implies that the signal posterior is Gaussian with mean $m=\langle s \rangle_{(s|d)}$ and covariance $D=\langle (s-m) \, (s-m)^\dagger \rangle_{(s|d)}$,
\begin{equation}
 \mathcal{P}(s|d)=\mathcal{G}(s-m,D),
\end{equation}
a result well known in Wiener filter theory of signal reconstruction \cite{1949wiener-short}. 

In a field theoretical language, the data dependent $j$ is an information source field, which excites our knowledge on $s$ being non-zero (as the preferred prior value was). The Wiener variance $D$ plays two distinct roles. On the one hand it is the susceptibility of our mean field $m$ to the force of the information source $j$, since $m=Dj$, on the other hand it describes the remaining a posteriori uncertainty  $D=\langle (s-m) \, (s-m)^\dagger \rangle_{(s|d)}$. In a field theoretical language, $D$ is the information propagator, since $D_{xy}$ transports the information source at location $y$ to the location $x$ of interest  in $m_x = (Dj)_x=\int dy\, D_{xy}\,j_y$.

In practice, one will use an iterative linear algebra method like the conjugate gradient method to solve numerically the equation $D^{-1}\,m=j$ for $m$ on a computer \cite{2008MNRAS.389..497K}.

\subsection{Interacting theory}
\subsubsection{Interaction Hamiltonian}

If any of the assumptions of our Wiener filter theory scenario is violated, in that the signal response is non-linear, the field or the noise is non-Gaussian, the noise variance depends on the signal, or the noise or signal covariances are unknown and have to be determined from the data itself, the resulting information Hamiltonian will contain anharmonic terms. These terms couple the different eigenmodes of the information propagator and lead to an interacting field theory. 
In many cases the Hamiltonian can be Taylor-Fr{\'e}chet expanded as
\begin{equation}
 H(d,s)= -\ln\mathcal{P}(d,s)=\underbrace{H_0-j^\dagger s + \frac{1}{2}\,s^\dagger D^{-1} s}_{H_\mathrm{free}} + \underbrace{ \sum_{i=3}^\infty \dotsint \!\! (dx_1\cdots  dx_i)\; \Lambda_{x_1\ldots x_i}^{(i)}\; s_{x_1}\cdots s_{x_i}}_{H_\mathrm{int}},
\end{equation}
and thereby split into a free ($H_\mathrm{free}$) and an interaction ($H_\mathrm{int}$) part. 
 Let us assume that the interaction terms are small. This can often be achieved, i.e., by shifting the field values to $s'=s-s_\mathrm{cl}$, where $s_\mathrm{cl}$ is the minimum of the Hamiltonian, the classical field, or in inference language, the maximum a posteriori estimator. Expanding $H(d,s')=H(d, s=s_\mathrm{cl}+s')$ around $s'=0$ then often ensures small interaction terms around the origin.
 
In this case, it is possible to expand the mean field value, or any other quantity of interest, around its free theory value. Since the terms of such an expansion can become numerous and complex, this is best done diagrammatically.

\subsubsection{Feynman diagrams}

Feynman diagrams provide a diagrammatical expansion to calculate perturbatively field expectation values. We are not explaining here how they work in detail, which for IFT is detailed in \cite{2009PhRvD..80j5005E}. We rather stress the important point that the main elements of the diagrams, the lines connecting source points and interaction vertices, are just an application of the propagator $D$. Since this could be done numerically for the free theory/Wiener filter case, we are already equipped with the necessary computational tools to calculate more complex diagrams. For example, the mean field of an interacting theory might be
\begin{eqnarray}\label{eq:simpleDiagram}
m &=& \langle s\rangle_{(s|d)} =
\begin{minipage}[c]{1.0\fgwidth}
\includegraphics[width=\fgwidth]{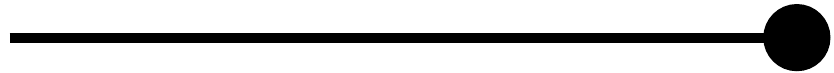} 
\end{minipage}
+\begin{minipage}[c]{1.0\fgwidth}
\includegraphics[width=\fgwidth]{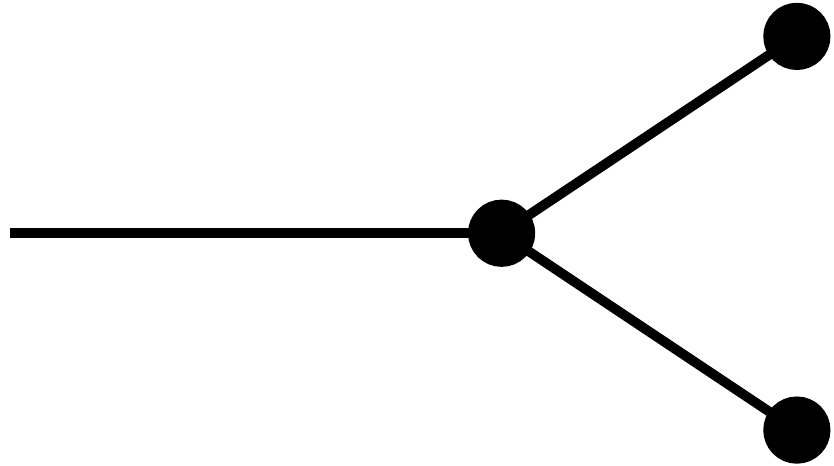} 
\end{minipage}
+\begin{minipage}[c]{1.0\fgwidth}
\includegraphics[width=\fgwidth]{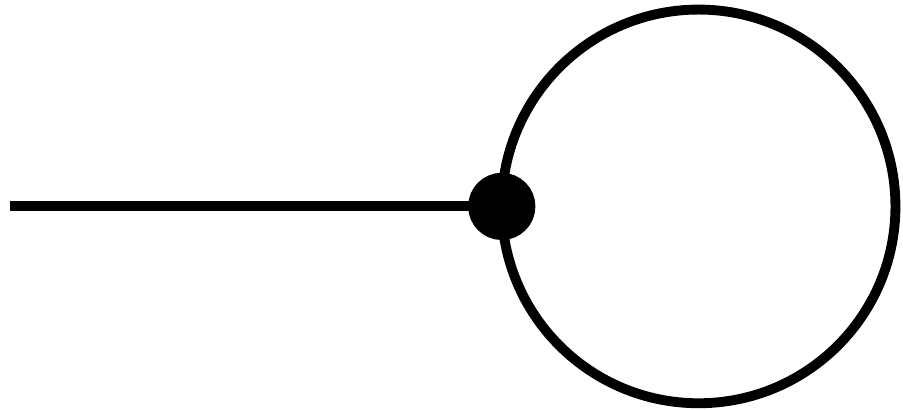} 
\end{minipage}
+ \ldots\nonumber \\
&=& D\,j -\frac{1}{2}\,D\, \Lambda^{(3)}\,[\cdot,D\,j, D\,j]   - \frac{1}{2}\,D\,\Lambda^{(3)}[\cdot, D]  + \ldots,
\end{eqnarray}
where we introduced $\Lambda^{(n)}[a,b,\ldots]=\dotsint \!\! (dx_1\cdots  dx_n)\; \Lambda_{x_1\ldots x_n}^{(n)}\; a_{x_1}\, b_{x_2}\,\cdots$ as a compact tensor notation.
The first diagram gives the Wiener filter signal reconstruction. In the second diagram, two Wiener filter maps are combined by the $\Lambda^{(3)}$-interaction, and then propagated to form the first non-linear correction to the Wiener filter. In the third diagram, the Wiener covariance replaces the two Wiener maps of the previous diagram, providing a correction due to the non-linearity effects on the uncertainty structure. More complex diagrams might also provide significant corrections, and have then to be calculated too. However, their computation can always be based on the linear Wiener filter case of the free theory, and is therefore possible.

\subsubsection{Thermodynamical inference}

A diagrammatic perturbation calculation leads to well performing algorithms in case the interaction terms are small. If they are large, resummation and renormalization techniques can be used and have proven to lead to well performing algorithms even for very non-linear measurement situations \cite{2009PhRvD..80j5005E} or in cases where the signal covariance has to be inferred as well from the data used for the signal reconstruction \cite{2011PhRvD..83j5014E}.
 
These techniques can be complex, and the meaning of the results is not necessarily intuitively understood. For the treatment of highly interacting quantum field theories, the effective action approach has proven helpful. The effective action is the Gibbs free energy $G$ known from thermodynamics  (here with temperature $T=1$), and this energy has the property that the map $m$, which minimizes it, is the desired mean field $m=\langle s\rangle_{(s|d)}$ given all constraints by the data.

The Gibbs free energy is the Legendre transformed Helmholtz free energy, which itself is (basically) the logarithm of the partition function $Z_d$. If we could calculate the partition function, we would be able to calculate mean field reconstruction directly from it via derivation with respect to the information source coefficient:
\begin{equation}
 \langle s\rangle_{(s|d)} = \frac{\delta \ln Z_d}{\delta j}.
\end{equation}
Thus, on a first sight, we did not win anything by reformulating the inference problem in terms of a Gibbs free energy, since this can only be calculated exactly in case we already have solved it.

However, the Gibbs free energy can also be expressed in terms of the internal energy $U=\langle H(d,s)\rangle_{(s|d)}=\int \mathcal{D}s\,\mathcal{P}(s|d)\, H(d,s)$ and the Boltzmann entropy $S_\mathrm{B}=-\int \mathcal{D}s\,\mathcal{P}(s|d)\, \ln \mathcal{P}(s|d)$ as
\begin{equation}
 G= U-T\,S_\mathrm{B}.
\end{equation}
This allows for a convenient approximative scheme, by replacing $\mathcal{P}(s|d)$ in the above definitions with an approximative Gaussian surrogate $\mathcal{G}(s-m,D)$ (except for the Hamiltonian in $U$), with mean $m$ and dispersion $D$ still to be determined.
This replacement turns the definitions for $U$ and $S_\mathrm{B}$ into Gaussian integrals, which can often be calculated analytically, e.g. $S_\mathrm{B}\approx \frac{1}{2} \mathrm{tr}(1+\ln(2\pi D))$.

Minimizing the resulting Gibbs free energy with respect to the unknown $m$ and $D$ gives then equations determining these quantities approximatively. This method of thermodynamical inference has proven to reproduce previously found results from renormalization and resummation calculations with much less effort \cite{2010PhRvE..82e1112E}. It was also very useful in developing novel algorithms, e.g. to deal with the problem of reconstructing a Gaussian signal field where 
the signal covariance is unknown but spectral smoothness can be assumed \cite{2012arXiv1210.6866O} or where
both the signal and the noise covariance where not known \cite{2011PhRvE..84d1118O}. The resulting algorithm, named extended critical filter, was successfully used for a reconstruction of the Galactic Faraday rotation sky signal \cite{2012A&A...542A..93O-short}.  

It is interesting to note that this minimal Gibbs free energy is equivalent to a minimal Kullback Leibler distance of $\mathcal{G}(s-m,D)$ to $\mathcal{P}(s|d)$ or to Maximum Entropy for $\mathcal{G}(s-m,D)$ with $\mathcal{P}(s|d)$ as the prior distribution \cite{2010PhRvE..82e1112E}. Thus information theory has basically reformulated methods developed earlier in thermodynamics, e.g. see \cite{2007AIPC..954...11C}. 


\section{Applications}

As the general theory of signal field inference, IFT has vast applications of which I want to mention a few listed at  \url{www.mpa-garching.mpg.de/ift}.

\textbf{{Cosmic magnetism}} studies have already been mentioned. IFT was here used to construct Galactic Faraday rotation maps from noisy data with unreliable noise information \cite{2012A&A...542A..93O-short}. The resulting maps can be analysed in order to test for helicity in Galactic magnetic fields \cite{2011A&A...530A..88J, 2011A&A...530A..89O}.

\textbf{{Cosmography}} is the 3-d cartography of the Cosmos. The main landmarks are the ambundant galaxies tracing the filamentary and knotty distribution of dark matter in space. Initial studies used Wiener filtering \cite{2008MNRAS.389..497K, 2009MNRAS.400..183K-short}, later the log-normal-Poisson model \cite{2009PhRvD..80j5005E, 2009arXiv0911.2496J, 2009arXiv0911.2498J, 2010MNRAS.403..589K, 2010MNRAS.409.1393W}, whereas the latest use the evolution of the Gaussian initial conditions into the observed density field  \cite{2012arXiv1203.3639J, 2012arXiv1203.4184K}.

\textbf{{Cosmic Microwave Background}} (CMB) studies are particularly well suited for IFT, since the CMB temperature statistics is very Gaussian. The weak non-Gaussianity is scientifically extremely interesting, since it is one of the few characteristic signatures of the inflationary epoch. An IFT data filter to search for such non-Gaussianity reproduces already known non-Gaussianity detection methods, while transfering them into a Bayesian setting  \cite{2009PhRvD..80j5005E}. Cross correlation studies of CMB and cosmic structure are also conveniently formulated in an IFT-language \cite{2008MNRAS.391.1315F, 2009MNRAS.395.1837F, 2010MNRAS.403.1739F}.

\textbf{Stochastic estimation} methods are widespread in numerics. For example, the diagonals and traces of complex numerical operators on high-dimensional function spaces (e.g. like the propagator $D$ of IFT) are often calculated
approximatively via stochastic probing. However, the real space structure of many such operator diagonals often exhibits sufficient smoothness that IFT methods can speed up their calculation \cite{2011arXiv1108.0600S}. 

\textbf{Numerical simulations} of partial differential equations face the problem that their differential operators require continuous fields to act on, but the data in computer memory is discrete. Thus a specific sub-grid field structure is usually assumed by conventional simulations schemes. IFT permits to construct the ensemble of plausible continuous fields being consistent with the data and other knowledge on which the operators can act in order to produce the time evolved field ensemble. A recast of this into an ensemble described by computer-data using entropic matching leads to a new and eventually better simulation methodology, called information field dynamics \cite{2012arXiv1206.4229E}.


\begin{theacknowledgments}
I want to thank all my students and coworkers, who accompanied me on the journey into the realm of IFT and gave me valuable guidance through feedback, discussions, and their own research. These are Michael Bell, Mona Frommert, Maksim Greiner, Jens Jasche, Henrik Junklewitz, Francisco Shu Kitaura, Niels Oppermann, Marco Selig, Maximilian Ullherr, Cornelius Weig, Helin Weingartner, and Lars Winderling. Further, I want to thank John Skilling and an anonymous referee for helpful comments on the manuscript.
\end{theacknowledgments}



\bibliographystyle{aipproc}   

\bibliography{../Bib/ift}

\end{document}